\documentclass[conference]{IEEEtran}
\usepackage[nolist]{acronym}
\usepackage{float}
\usepackage{fontawesome}
\usepackage{amsmath}
\usepackage{tikz}

\usepackage{graphicx}
\usepackage{subcaption}
\usepackage{xcolor}     
\usepackage{tabularray}  
\let\labelindent\relax
\usepackage{enumitem}
\usepackage[T1]{fontenc}
\usepackage{url}

\ifCLASSINFOpdf
\else
\fi
\begin{document}
\acrodef{3GPP}{3rd Generation Partnership Project}
\acrodef{5G}{5th Generation Mobile Network}
\acrodef{6G}{6th Generation Mobile Network}
\acrodef{AI}{Artificial Intelligence}
\acrodef{AI4Net}{\ac{AI} for Networking}
\acrodef{AIDER}{Aerial Image Dataset for Emergency Response}
\acrodef{AIaaS}{Artificial Intelligence-as-a-Service}
\acrodef{AMF}{Access and Mobility Management Function}
\acrodef{API}{Application Programming Interface}
\acrodef{AUSF}{Authentication Server Function}
\acrodef{IID}{Independent and Identically Distributed}
\acrodef{B5G}{Beyond Fifth Generation}
\acrodef{BPF}{Berkeley Packet Filter}
\acrodef{CA}{Certificate Authority}
\acrodef{CBR}{Constant Bit Rate}
\acrodef{CI}{Confidence Interval}
\acrodef{CV}{Coefficient of Variation}
\acrodef{CSV}{Comma-Separated Values}
\acrodef{CPU}{Central Processing Unit}
\acrodef{CNN}{Convolutional Neural Network}
\acrodef{CNNs}{Convolutional Neural Networks}
\acrodef{CRQC}{Cryptographically Relevant Quantum Computer}

\acrodef{DCA}{Dynamic Cryptographic Agility}
\acrodef{DoS}{Denial of Service}
\acrodef{DDoS}{Distributed Denial of Service}
\acrodef{DDPG}{Deep Deterministic Policy Gradient}
\acrodef{DNN}{Deep Neural Network}
\acrodef{DSA}{Digital Signature Algorithms}
\acrodef{DRL}{Deep Reinforcement Learning}
\acrodef{DT}{Decision Tree}

\acrodef{ETSI}{European Telecommunications Standards Institute}
\acrodef{eNWDAF}{Evolved Network Data Analytics Function}
\acrodef{ECC}{Elliptic-curve cryptography}
\acrodef{E2E}{End-To-End}

\acrodef{FIBRE}{Future Internet Brazilian Environment for Experimentation}
\acrodef{Fuzz}{(Fuzzing or Fuzz testing) - Automated Software Testing Technique}

\acrodef{GNN}{Graph Neural Networks}
\acrodef{GPU}{Graphics Processing Unit}
\acrodef{HTM}{Hierarchical Temporal Memory}
\acrodef{HNDL}{Harvest Now, Decrypt Later}
\acrodef{HTTP}{Hypertext Transfer Protocol}
\acrodef{HTTPS}{Hypertext Transfer Protocol Security}
\acrodef{HSM}{Hardware Security Module}

\acrodef{IAM}{Identity And Access Management}
\acrodef{ICMP}{Internet Control Message Protocol}
\acrodef{IID}{Informally, Identically Distributed}
\acrodef{IoE}{Internet of Everything}
\acrodef{IoT}{Internet of Things}
\acrodef{IQR}{Interquartile Range}
\acrodef{IPsec}{Internet Protocol Security}
\acrodef{ITU}{International Telecommunication Union}
\acrodef{KEM}{Key Encapsulation Mechanism}
\acrodef{KNN}{K-Nearest Neighbors}
\acrodef{KPI}{Key Performance Indicator}
\acrodef{KPIs}{Key Performance Indicators}
\acrodef{LSTM}{Long Short-Term Memory}
\acrodef{LTS}{Long Term Support }

\acrodef{MITM}{Man-in-the-Middle}
\acrodef{MAE}{Mean Absolute Error}
\acrodef{MACsec}{Media Access Control Security}
\acrodef{ML}{Machine Learning}
\acrodef{MLaaS}{Machine Learning as a Service}
\acrodef{MOS}{Mean Opinion Score}
\acrodef{MAPE}{Mean Absolute Percentage Error}
\acrodef{MSE}{Mean Squared Error}
\acrodef{MEC}{Multi-access Edge Computing}
\acrodef{mMTC}{Massive Machine Type Communications}
\acrodef{MFA}{Multi-factor Authentication}

\acrodef{NWDAF}{Network Data Analytics Function}
\acrodef{NSSF}{Network Slice Selection Function}
\acrodef{Net4AI}{Networking for \ac{AI}}
\acrodef{NIST}{National Institute of Standards and Technology}
\acrodef{NS}{Network Slicing}
\acrodef{NF}{Network Function}
\acrodef{NFV}{Network Function Virtualization}
\acrodef{NFVM}{Network Function Virtualization Management}
\acrodef{NRF}{Network Repository Function}
  
\acrodef{OSM}{Open Source MANO}
\acrodef{OAuth}{Open Authorization}
\acrodef{OQS}{Open Quantum Safe}

\acrodef{PCA}{Principal Component Analysis}
\acrodef{PDU}{Packet Data Unit}
\acrodef{PCF}{Policy Control Function}
\acrodef{PoC}{Proof of Concept}
\acrodef{PQC}{Post-Quantum Cryptography }

\acrodef{QoE}{Quality of experience}
\acrodef{QoS}{Quality of Service}

\acrodef{RAM}{Random Access Memory}
\acrodef{RESTful}{REpresentational State Transfer Application Programming Interface}
\acrodef{RF}{Random Forest}
\acrodef{RL}{Reinforcement Learning}
\acrodef{RMSE}{Root Mean Square Error}
\acrodef{RNN}{Recurrent Neural Network}
\acrodef{RTT}{Round-Trip Time}
\acrodef{RAN}{Radio Access Network}
\acrodef{RSA}{Rivest–Shamir–Adleman}
\acrodef{RX}{Receive}
\acrodef{SBA}{Service-Based Architecture}
\acrodef{SBI}{Service-Based Interface}
\acrodef{SCTP}{Stream Control Transmission Protocol}
\acrodef{SDN}{Software-Defined Networking}
\acrodef{SFI2}{Slicing Future Internet Infrastructures}
\acrodef{SLA}{Service-Level Agreement}
\acrodef{SON}{Self-Organizing Network}
\acrodef{SMF}{Session Management Function}
\acrodef{S-NSSAI}{Single Network Slice Selection Assistance Information}
\acrodef{SPPQC}{Sidecar Proxy \ac{PQC}}
\acrodef{SVM}{Support Vector Machine}

\acrodef{TCP}{Transmission Control Protocol}
\acrodef{TEE}{Trusted Execution Environment}
\acrodef{TQFL}{Trust Deep Q-learning Federated Learning}
\acrodef{TLS}{Transport Layer Security}
\acrodef{TLSSP}{\ac{TLS} Sidecar Proxy}
\acrodef{TEID}{Tunnel Endpoint Identifier}
\acrodef{TX}{Transmit}

\acrodef{UE}{User Equipment}
\acrodef{UDM}{Unified Data Management}
\acrodef{UPF}{User Plane Function}
\acrodef{UDR}{Unified Data Repository}
\acrodef{URLLC}{Ultra-Reliable and Low Latency Communications}
\acrodef{UAV}{Unmanned Aerial Vehicle}
\acrodef{UAVs}{Unmanned Aerial Vehicles}

\acrodef{VoD}{Video on Demand}
\acrodef{VM}{Virtual Machine}
\acrodef{VR}{Virtual Reality}
\acrodef{AR}{Augmented Reality}
\acrodef{V2V}{Vehicle-to-Vechile}
\acrodef{V2X}{Vehicle-to-Everything}

%


\title{Empowering Mobile Networks Security Resilience by using Post-Quantum Cryptography}

\author{\IEEEauthorblockN{
Ricardo Alves Faval\textsuperscript{1},
Rodrigo Moreira\textsuperscript{2}, 
Flávio de Oliveira Silva\textsuperscript{1,}\textsuperscript{3}}
\IEEEauthorblockA{
\textsuperscript{1}Federal University of Uberlândia (UFU), Minas Gerais, Brazil\\
\textsuperscript{2}Federal University of Viçosa (UFV), Minas Gerais, Brazil\\
\textsuperscript{3}University of Minho (UMinho), Braga, Portugal\\
Emails: ricardo.faval@ufu.br, rodrigo@ufv.br, flavio@di.uminho.pt}
}


%


\maketitle

\begin{abstract}
The transition to a cloud-native 5G Service-Based Architecture (SBA) improves scalability but exposes control-plane signaling to emerging quantum threats, including Harvest-Now, Decrypt-Later (HNDL) attacks. While NIST has standardized post-quantum cryptography (PQC), practical, deployable integration in operational 5G cores remains underexplored. This work experimentally integrates NIST-standardized ML-KEM-768 and ML-DSA into an open-source 5G core (free5GC) using a sidecar proxy pattern that preserves unmodified network functions (NFs). Implemented on free5GC, we compare three deployments: (i) native HTTPS/TLS, (ii) TLS sidecar, and (iii) PQC-enabled sidecar. Measurements at the HTTP/2 request--response boundary over repeated independent runs show that PQC increases end-to-end Service-Based Interface (SBI) latency to $\sim$54~ms, adding a deterministic 48--49~ms overhead relative to the classical baseline, while maintaining tightly bounded variance ($IQR \leq 0.2$~ms, $CV < 0.4\%$). We also quantify the impact of Certification Authority (CA) security levels, identifying certificate validation as a tunable contributor to overall delay. Overall, the results demonstrate that sidecar-based PQC insertion enables a non-disruptive and operationally predictable migration path for quantum-resilient 5G signaling.
\end{abstract}

\IEEEpeerreviewmaketitle

\section{Introduction}\label{sec:introduction}

The deployment of \ac{5G} networks represents a paradigm shift in telecommunications, supporting a massive ecosystem of billions of devices. Central to this evolution is the \ac{SBA}, which replaced traditional point-to-point interfaces with a cloud-native, microservices-oriented approach. While this transition to \ac{HTTP}/2 and \ac{TLS} enhances scalability, it also centralizes sensitive cryptographic operations at critical core network functions \acp{NF}~\cite{Zhou2023, Hoque2025}.

As research progresses toward \ac{6G} networks, the security foundations of these infrastructures face an existential threat: the advancement of quantum computing. Current \ac{5G} security relies heavily on public-key cryptography, specifically \ac{RSA} and \ac{ECC}, which are vulnerable to \acp{CRQC}. Attacks such as \acf{HNDL} \cite{sanon_securing_2025} underscore the urgency, as adversaries may already be capturing sensitive \ac{5G} signaling to decrypt once powerful quantum machines become available. This poses a long-term risk to subscriber identities and authentication credentials stored in \ac{UDM} modules~\cite{Scalise2024, Hoque2025}. 

To mitigate these risks, the \ac{NIST} finalized the first Post-Quantum Cryptography \ac{PQC} standards in 2024, emphasizing \ac{KEM} and \ac{DSA}. However, a significant gap remains between theoretical standardization and practical implementation in high-throughput environments like the \ac{5G} Core~\cite{Olushola2026, Truong2026}.

This work addresses this gap by presenting a comprehensive evaluation of \ac{PQC} integration into a cloud-native \ac{5G} Core. We propose a sidecar proxy architecture that enables quantum-safe communication between \acp{NF} without modifying the legacy codebase. Specifically, we integrate ML-KEM\cite{Chuanming2025a} and ML-DSA\cite{ASTARLOA2025} into the free5GC platform. Our main contribution is a rigorous experimental comparison across three scenarios: (i) a baseline Native\ac{5G} Core; (ii) a \ac{TLS}-based sidecar; and (iii) our \ac{PQC}-enabled sidecar architecture. We provide detailed insights into the impact of larger \ac{PQC} keys on control plane latency, handshake performance, and resource consumption (\ac{CPU} /Memory), offering a roadmap for future quantum-resistant telecommunications infrastructure.

To the best of our knowledge, this work represents one of the first fully experimental integrations of NIST-standardized \ac{PQC} algorithms into a cloud-native \ac{5G} Core using a sidecar proxy pattern that preserves unmodified \ac{NF} implementations. Unlike prior studies that focus on theoretical security analysis, protocol-level proposals, or hybrid cryptographic simulations, our approach provides a deployable migration strategy within a production-grade free5GC environment. 

Beyond reporting end-to-end latency, we experimentally isolate cryptographic processing overhead from wrapper-induced proxy costs, quantify deterministic control-plane behavior under \ac{PQC}, and evaluate the impact of different \ac{CA} security parameter levels. This enables a practical characterization of performance–security trade-offs and demonstrates crypto-agility in \ac{SBA} signaling without intrusive architectural changes.


The remainder of this paper is organized as follows. Section~\ref{sec:related_work} discusses related work on \ac{PQC} and its adoption in \ac{5G} and cloud-native deployments. Section~\ref{sec:threat_model} presents the threat model and security goals. Section~\ref{sec:proposed_method} describes the proposed sidecar-based architecture for protecting SBI and extending coverage to N2 signaling. Section ~\ref{sec:pqc_enabled_free5gc} details the implementation to empower Free5GC with \ac{PQC}. Section~\ref{sec:experimental_setup} presents the experimental methodology, including evaluation scenarios and measurement approach. Section~\ref{sec:results_and_dicussion} reports and analyzes the results, and Section~\ref{sec:conclusion} concludes the paper and outlines future work.

\section{Related Work}\label{sec:related_work}

\begin{table*}[t]
\centering
\caption{Short State of the art survey.}
\label{tab:sota_survey}
\resizebox{\textwidth}{!}{
\begin{tabular}{lccccccccccc}
\hline
Paper & 5GC/SBA & TLS & IPsec & OAuth & Exp. & 5G Stack & API/Fuzz & \ac{PQC} \\
\hline
Mahyoub et al.~\cite{Mahyoub2024} &  \faCircle &  \faCircle &  \faCircle &  \faCircle &  \faCircleO &    \faCircleO &  \faCircleO &  \faCircle \\
Mehic et al.~\cite{Mehic2024} &  \faCircle &  \faCircle &  \faCircle &  \faCircleO &  \faCircleO &   \faCircleO &  \faCircleO &  \faCircle \\
Lawo et al.~\cite{Lawo2024} &  \faCircleO &  \faCircleO &  \faCircleO &  \faCircleO &  \faCircleO &    \faCircleO &  \faCircleO &  \faCircle \\
Scalise et al.~\cite{Scalise2024} &  \faCircle &  \faCircle &  \faCircle &  \faCircle &  \faCircleO &   \faCircleO &  \faCircleO &  \faCircle \\
Mangla et al.~\cite{Mangla2023} &  \faCircleO &  \faCircle &  \faCircle &  \faCircleO &  \faCircleO &    \faCircleO &  \faCircleO &  \faCircle \\
Dolente et al.~\cite{Dolente2024} &  \faCircle &  \faCircle &  \faCircle &  \faCircle &  \faCircleO &    \faCircle &  \faCircle &  \faCircleO \\
Pell et al.~\cite{Pell2023} &  \faCircle &  \faCircleO &  \faCircleO &  \faCircleO &  \faCircle &  \faCircle &  \faCircleO &  \faCircleO \\
\textbf{This work} & \faCircle  &  \faCircle &  \faCircle &  \faCircle &  \faCircle &  \faCircle &  \faCircle &  \faCircle \\
\hline
\end{tabular}}
\end{table*}

Mahyoub et al.~\cite{Mahyoub2024} present a security analysis of critical interfaces in the Fifth Generation architecture, identifying interfaces that exchange sensitive data or are externally exposed, and mapping threats and protections across Service-Based Architecture and non-Service-Based Architecture connectivity. It discusses defenses such as Transport Layer Security, Internet Protocol Security, and Open Authorization for Network Function interactions, and highlights post-quantum considerations for key exchange and tunnel establishment.

Mehic et al.~\cite{Mehic2024} provide a comprehensive overview of quantum cryptography for Fifth Generation networks, covering Quantum Key Distribution deployment options, integration challenges, and standardization aspects in cellular architectures. It contrasts Quantum Key Distribution with \ac{PQC} and discusses how quantum-resilient key establishment can complement mechanisms such as \ac{TLS} and \ac{IPsec} in Fifth Generation systems.

Lawo et al.~\cite{Lawo2024} designs and implements a Post Quantum Cryptography network stack on a Data Processing Unit platform, combining Falcon with Kyber and Dilithium with Kyber for authentication and key establishment. It benchmarks performance on the NVIDIA Data Processing Unit platform, reporting latency and throughput trade-offs for \ac{PQC}-protected communication.


Scalise et al.~\cite{Scalise2024} offer a systematic survey of security considerations in Fifth and Sixth Generation systems, organizing challenges and trends across confidentiality, integrity, and availability objectives, and introducing a Zero Trust Architecture perspective. It highlights emerging directions, including Post-Quantum Cryptography for key exchange, as well as the security implications of Software-Defined Networking, Network Function Virtualization, and Multi-access Edge Computing deployments.

Mangla et al.~\cite{Mangla2023} review Fifth Generation security vulnerabilities in a future quantum computing environment and survey quantum-based solutions intended to mitigate these challenges, while motivating the transition toward Sixth Generation. It discusses Quantum Key Distribution and related quantum security techniques and argues that classical public key cryptography must be complemented by quantum-resilient alternatives for next-generation industrial deployments.


Dolente et al.~\cite{Dolente2024} perform an experimental vulnerability assessment of open-source Fifth Generation Core Network Function implementations, focusing on externally exposed Network Functions such as the Access and Mobility Management Function and the Network Repository Function with Network Exposure Function. It applies Application Programming Interface injection and fuzzing-style tests to Open5GS and OpenAirInterface, and reports robustness issues that can lead to denial-of-service attacks and other security risks.

Pell et al.~\cite{Pell2023} study service classification of Fifth Generation core network traffic using Machine Learning over flow and packet metadata, motivated by detecting malicious signalling without Deep Packet Inspection. It compares multiple Machine Learning models and shows how accurate service-level inference can support security monitoring for distributed Network Functions in Fifth-Generation core networks.

In Table~\ref{tab:sota_survey}, \ac{5G} / \ac{SBA} indicates an explicit focus on the \ac{6G} and \ac{SBI} and \ac{TLS}, \ac{IPsec}, and \ac{OAuth} indicate whether \ac{TLS}, \ac{IPsec}, and \ac{OAuth} are discussed or used. \ac{5G} Stack indicates use of an open-source Fifth Generation implementation or testbed; \ac{API}/\ac{Fuzz} indicates \ac{API} injection or fuzzing-style testing; and \ac{PQC} indicates explicit \ac{PQC} discussion or implementation.

\section{Threat Model and Security Goals} \label{sec:threat_model}

The threat model considers a passive adversary capable of recording encrypted traffic for later decryption in a \ac{HNDL} scenario, as well as an active adversary performing \ac{MITM} attacks to intercept, replay, or modify signaling data. We assume that the \acp{NF} and their host environments are trusted, while the underlying transport network is inherently insecure. While trusted \ac{NF} execution is assumed, integration with \acp{TEE} or confidential-computing mechanisms represents an important direction for mitigating endpoint compromise in future deployments.

The protected assets are the confidentiality and integrity of \ac{SBI} control-plane signaling exchanged between \acp{NF} (e.g., registration, discovery, and heartbeat) over potentially untrusted transport links. Consequently, the primary security objectives are to ensure long-term confidentiality, robust mutual authentication, and message integrity through post-quantum mechanisms. Furthermore, the architecture must support crypto-agility, enabling an incremental migration to \ac{PQC} without requiring modifications to the internal logic of the \acp{NF}. 

This model aligns with \ac{3GPP} assumptions where protection is applied at the \ac{SBA} level; however, we reinforce these interfaces against quantum-capable adversaries. Out-of-scope threats include endpoint compromise, \ac{DoS} attacks, side-channel analysis, and radio-access security.

By design, the sidecar introduces an additional trusted component in the \ac{NF} pod, so reducing its attack surface (minimal codebase, least-privilege access, and hardened container configuration) is a key operational requirement.

\section{Empowering the Core \ac{5G} with \ac{PQC}}\label{sec:proposed_method}
The transition to \ac{PQC} is one of the most important steps for telecommunications security, but the direct integration into the network ecosystem often entails complex modifications to legacy software, resulting in compatibility issues and increased maintenance costs. 
To overcome these challenges, this work proposes a  sidecar proxy pattern~\cite{aldas_cloud-native_2023} that decouples security functions from the core business logic of the \ac{NF}. Instead of modifying the \ac{NF} source code, specialized proxies are deployed alongside each component to handle \ac{PQC}-protected handshakes and key exchanges transparently. This modular layer abstracts the complexities of quantum-resistant protocols, creating a "plug-and-play" environment where algorithms can be updated without disrupting the underlying \ac{5G} infrastructure.

For this implementation, the ML-\ac{KEM}-768 and ML-\ac{DSA}-65 algorithms were selected due to their optimal balance between post-quantum resilience and operational efficiency, aligning with \ac{NIST} Security Level 3 standards~\cite{alagic2025status}. While higher-level parameters (e.g., Level 5) offer superior security margins, they introduce prohibitive communication overhead and handshake latency \cite{MONTENEGRO2026111957} that could compromise the strict performance requirements of the \ac{5G} \ac{SBA}. Furthermore, these specific parameter sets benefit from mature implementations and extensive optimizations in digital signature calculations and verification processes~\cite{Gorbenko_Kachko_Derevianko_2025}. By integrating these standardized algorithms via the sidecar approach, the architecture ensures robust, quantum-resistant protection while maintaining the high-performance throughput and low-latency response times essential for mission-critical core network operations.

While the architecture supports \ac{PQC} encapsulation for N2 signaling, the experimental evaluation presented in this work focuses exclusively on \ac{SBI} interactions to isolate control-plane cryptographic overhead.

\section{The \ac{PQC} Enabled Free5GC}\label{sec:pqc_enabled_free5gc}

\begin{figure}[htbp]
\centering
\includegraphics[width=.48\textwidth]{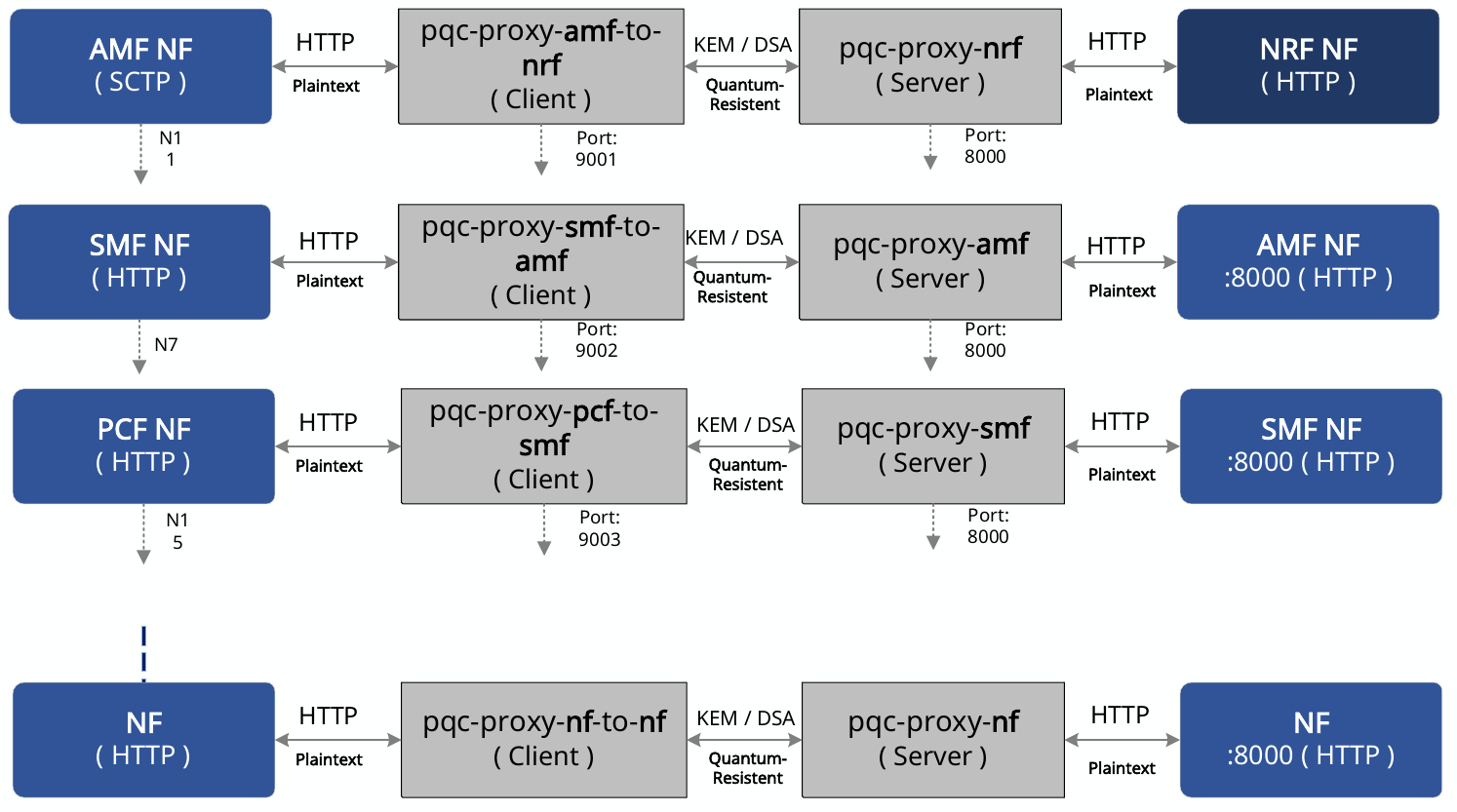}
\caption{PQC Secure Free5GC Architecture (partial view)}
\label{fig:pqc_free5gc_architecture}
\end{figure}

\begin{figure}[H]
\centering
\includegraphics[width=.48\textwidth]{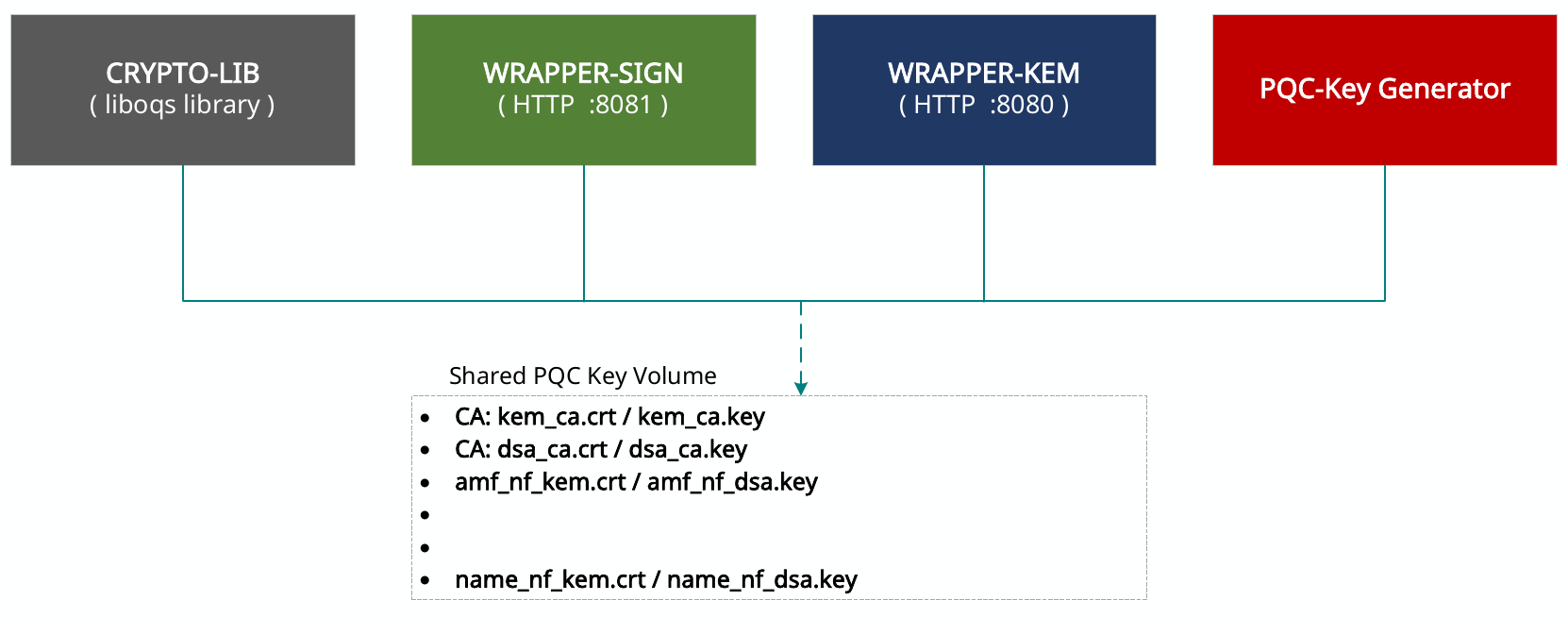}
\caption{Cryptographic infrastructure (partial view).}
\label{fig:pqc_infrastructure_layer}
\end{figure}

The experimental environment was built upon the free5GC platform, an open-source implementation of \ac{3GPP} \ac{5G} Core specifications designed for cloud-native deployments. Utilizing the free5gc-compose project, we implemented a modular post-quantum infrastructure using a sidecar proxy pattern that strictly separates cryptographic operations from the \ac{NF} business logic. As illustrated in the architecture (see Figure ~\ref{fig:pqc_free5gc_architecture}), the deployment relies on a specialized cryptolib container integrating liboqs v0.15.0 and the OpenSSL \ac{OQS}-provider, establishing the foundation for quantum-resistant primitives. To abstract these low-level interactions, we developed wrapper-kem and wrapper-sign services that expose \ac{RESTful} \acp{API} for ML-\ac{KEM} and ML-\ac{DSA} operations. This structure (see Figure ~\ref{fig:pqc_infrastructure_layer}), orchestrated via Docker Compose, ensures that cryptographic lifecycles, including the generation of entity-specific keys and certificates by a dedicated \ac{PQC}-key-generator are managed independently of the \ac{NF}.

\begin{figure}[htbp]
\centering
\includegraphics[width=.48\textwidth]{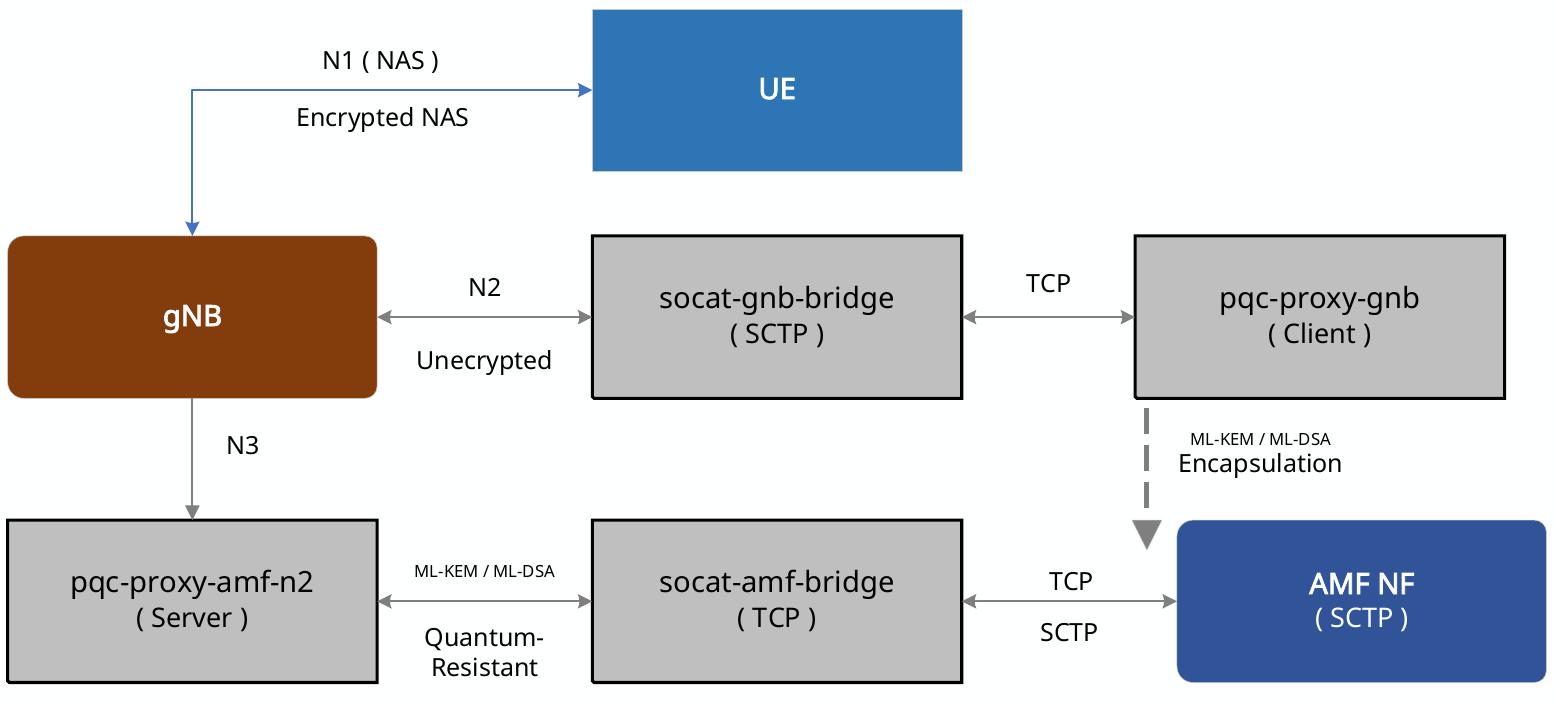}
\caption{\ac{PQC} \ac{RAN} domain (partial view).}
\label{fig:pqc_ran_domain}
\end{figure}

The core of the \ac{PQC} integration lies in the pqc-proxy instances deployed in both server and client modes. \acp{NF} communicate via plain-text \ac{HTTP} with local proxies within isolated Docker bridge networks (172.17.0.0/16), while inter-proxy traffic across network edges is encapsulated in ML-\ac{KEM}-768 tunnels and authenticated via ML-\ac{DSA}-65. A critical technical challenge addressed was the protection of the N2 interface; since NGAP signaling typically relies on \ac{SCTP}, we deployed socat bridges to perform \ac{SCTP}-to-\ac{TCP} conversion, enabling \ac{PQC} encapsulation between the pqc-proxy-gnb and pqc-proxy-amf-n2 (see Figure ~\ref{fig:pqc_ran_domain}). To ensure traffic redirection, the free5GC configuration files (e.g., amfcfg.yaml, nrfcfg.yaml) were modified to route \ac{SBI} requests through the local proxy’s listener port.

\begin{figure}[htbp]
\centering
\includegraphics[width=.48\textwidth]{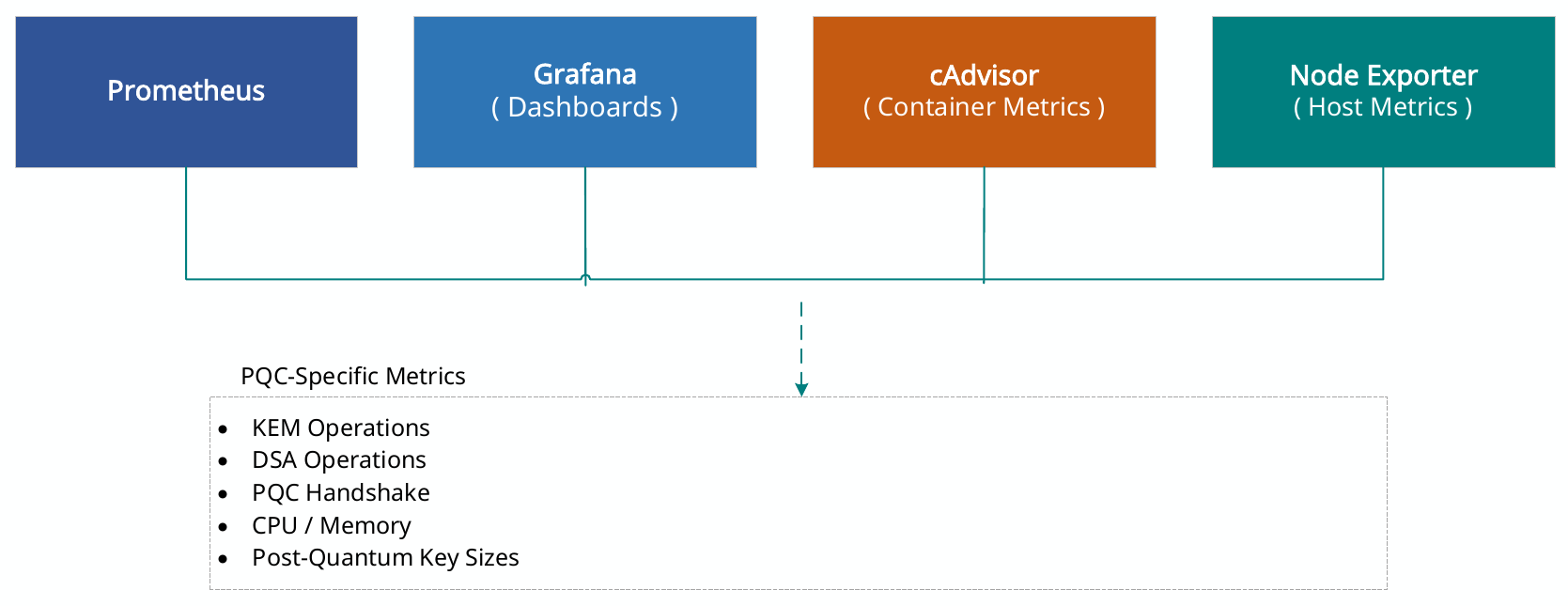}
\caption{Monitoring infrastructure layer.}
\label{fig:pqc_monitoring_layer}
\end{figure}

To evaluate the resulting cryptographic overhead and system stability, a robust monitoring framework was integrated directly into the architecture. This stack comprises Prometheus for metrics collection, Grafana for real-time visualization, Node Exporter for hardware telemetry, and cAdvisor for container-level resource tracking. This integrated observability framework enables precise, high-fidelity analysis of \ac{CPU} and memory consumption, as well as latency variations across the \ac{PQC} mechanisms deployed within the \ac{5G} Core, providing the necessary data to validate the performance trade-offs of the proposed sidecar approach (see Figure \ref{fig:pqc_monitoring_layer}).

\section{Experimental Evaluation}
\label{sec:experimental_setup}

This section presents the experimental methodology adopted to assess the performance implications of integrating \ac{PQC} into the free5GC using the proposed sidecar architecture. The evaluation is designed to ensure reproducibility, isolate cryptographic overhead from wrapper-induced processing costs, and enable a controlled comparison between Native, classical \ac{TLS}, and \ac{PQC}-enabled deployments. The following subsections describe the workload definition, testbed configuration, experimental design, and statistical reporting framework.

\subsection{Evaluation Scope and Workload}

To quantify the performance impact of the sidecar architecture and \ac{PQC}, the evaluation focuses on control-plane \ac{SBI} interactions between \acp{NF} and the \ac{NRF}, including registration, discovery, and heartbeat procedures. By isolating the control plane from \ac{UE} emulation overhead, the setup enables direct comparison between the Native architecture, a classical \ac{TLS} sidecar proxy, and a \ac{PQC} sidecar proxy integrated via liboqs.

The workload consists of sequential \ac{HTTP}/2 request--response exchanges per \ac{NF} pair. Each experimental repetition executes 100 transactions without artificial concurrency to isolate cryptographic overhead from scheduling or load-induced effects. A transaction is considered successful when the complete exchange finishes without timeout or retransmission.

\subsection{Testbed Configuration}

All architectures were deployed in standardized \acp{VM} running Ubuntu 22.04 \ac{LTS}, each allocated 4 \ac{CPU} cores and 8~GB of \ac{RAM}. The \ac{5G} Core was orchestrated using free5GC-compose v4.2.0. The \ac{PQC} implementation relied on OpenSSL v3.4.0 integrated with liboqs v0.15.0, enabling ML-\ac{KEM}-768 and ML-\ac{DSA} through sidecar proxies. All experiments were conducted without hardware acceleration, isolating software-side overheads.

\subsection{Experimental Design}

For each architecture (Native, classical \ac{TLS} Proxy, and \ac{PQC} Proxy), experiments were repeated 30 times, each repetition involving full container lifecycle reinitialization to capture variability from virtualized resource allocation and network stack initialization. Within each repetition, 100 sequential \ac{SBI} interactions were executed. The \ac{PQC} configuration was evaluated across three \ac{CA} parameter sets (ML-\ac{DSA}-44, 65, and 87), while maintaining ML-\ac{KEM}-768 and ML-\ac{DSA}-65 for key exchange and signatures.

\subsection{Metrics and Statistical Reporting}

Metrics were collected using Prometheus, cAdvisor, and Node Exporter. For statistical reporting, each repetition is treated as an independent sample. Results are expressed using median and \ac{IQR} across repetitions, and 95\% confidence intervals were computed over per-run median latency values using the $t$-distribution. The monitored performance and security indicators are summarized in Table~\ref{tbl:tbl_metrics_used}.

\definecolor{Matisse}{rgb}{0.098,0.407,0.643}
\begin{table}[htbp]
\centering
\caption{Selected metrics for evaluation.} 
\begin{tblr}{
  width = \linewidth,
  colspec = {X[1]X[1.5]X[3]},
  row{1} = {Matisse,fg=white,c,m},
  row{1} = {c},
  row{2-6} = {m},
  hlines, 
  vlines,
}
\textbf{Visual Analysis} & \textbf{Selected Metrics} & \textbf{Evaluation Goal} \\
Latency Distribution & $L_{SBI}$ & Assess delay scaling and jitter across architectures. \\
Resource Efficiency & { \ac{CPU}\% \\ $Mem_{MB}$ }& Validate hardware sustainability for \ac{PQC} workloads. \\
Latency Breakdown & { $L_{Srv}$ \\ $L_{Cli}$ } & Isolate sidecar overhead from cryptographic processing. \\
\ac{SBI} Matrix & $L_{SBI}$ per \ac{NF} pair & Identify bottlenecks in \ac{5G} signaling (e.g., \ac{AMF}, \ac{NRF}). \\
\ac{PQC} Operations & { $T_{enc}$\\ $T_{sig}$ \\ $T_{ver}$ }& Correlate KEM/DSA complexity with \ac{E2E} performance. \\
\end{tblr}
\label{tbl:tbl_metrics_used} 
\end{table}

\section{Results and Discussion}\label{sec:results_and_dicussion}

We evaluated three free5GC deployments: (i) \textit{Native}, using standard \ac{HTTPS}/\ac{TLS}; (ii) \textit{\ac{TLS} Proxy}, using a sidecar model with OpenSSL certificates; and (iii) \textit{\ac{PQC} Proxy}, enabling \ac{PQC} through the proposed sidecar approach.

\begin{figure}[htbp]
\centering
\includegraphics[width=.48\textwidth]{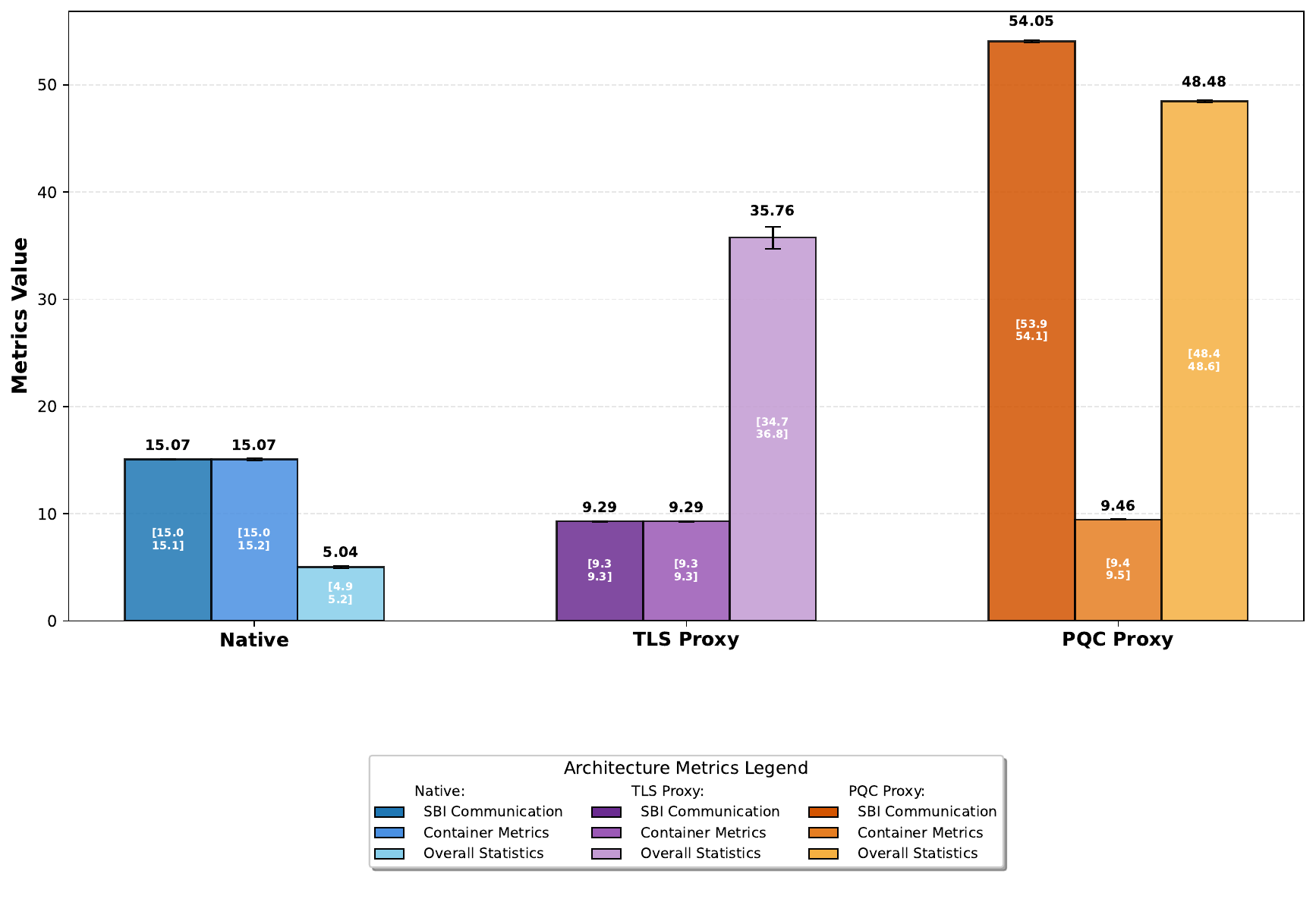}
\caption{Metrics comparison across architectures Core \ac{5G}.}
\label{fig:combined_metrics_comparison}
\end{figure}

Latency is measured at the \ac{HTTP}/2 request--response boundary on the \ac{SBI}: timestamps are captured immediately before request transmission and after complete response reception at the application layer. The \ac{SBI} Communication metric therefore captures end-to-end control-plane transaction latency for one \ac{NF}-to-\ac{NF} interaction. Container Metrics capture proxy-side processing behavior, while Overall Statistics correspond to the per-run aggregate of all collected \ac{SBI} request--response latencies across monitored \ac{NF} pairs (Figure~\ref{fig:combined_metrics_comparison}). Per-\ac{NF}-pair heatmaps report mean latencies for individual paths (e.g., 5--6~ms in the Native case), whereas Figure~\ref{fig:combined_metrics_comparison} summarizes distributional medians aggregated across interactions within each run.

For statistical reporting, 95\% confidence intervals were computed over per-run median latency values using the $t$-distribution across 30 independent repetitions, while error bars represent the \ac{IQR} ($Q1$--$Q3$) around the median. Native establishes the baseline, with \ac{SBI} Communication and Container Metrics medians of 15.07 (\ac{CI}: $[15.0, 15.1]$ and $[15.0, 15.2]$) and minimal quartile deviation ($\pm0.05$--$0.1$). Overall Statistics records 5.04 (\ac{CI}: $[4.9, 5.2]$), indicating stable conventional \ac{TLS} behavior.

The \ac{TLS} Proxy converges \ac{SBI} Communication and Container Metrics at 10.20~ms (\ac{CI}: $[10.2, 10.2]$ for both), corresponding to a 32.3\% reduction relative to Native and near-zero \ac{IQR}. This reduction does not imply lower cryptographic complexity; it is consistent with the sidecar centralizing connection management and reusing pooled \ac{HTTP}/2/\ac{TLS} sessions across repeated \ac{SBI} exchanges, thereby amortizing handshake and transport setup costs. To ensure comparability, latency was measured at the same request--response boundary in all architectures, using identical keep-alive and connection-reuse settings across runs.

\begin{figure}[htbp]
\centering
\includegraphics[width=.48\textwidth]{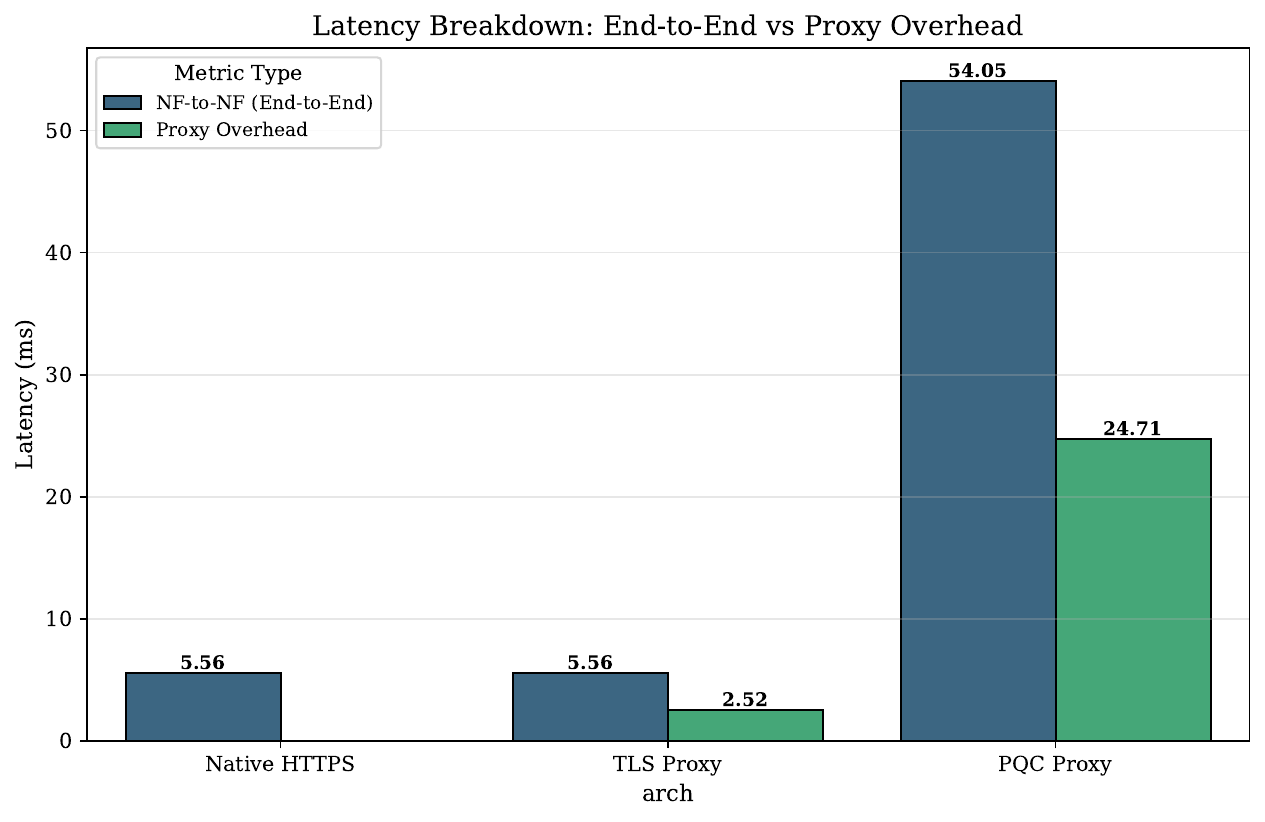}
\caption{Processing latency End-to-End.}
\label{fig:latency_breakdown_full}
\end{figure}

In contrast, the \ac{PQC} Proxy increases \ac{SBI} Communication to 54.05~ms (\ac{CI}: $[53.9, 54.1]$) while preserving tight quartile bounds ($\pm0.1$), quantifying a stable overhead of approximately 48--49~ms relative to classical \ac{TLS}. Container Metrics remain stable at 9.46~ms (\ac{CI}: $[9.4, 9.5]$), suggesting bounded resource behavior even under ML-\ac{KEM} and ML-\ac{DSA} processing (Figure~\ref{fig:latency_breakdown_full}).

From a standards perspective, the measured $\sim 54$~ms end-to-end \ac{SBI} latency remains well within \ac{5G} control-plane timing tolerances, which are governed by NAS procedure timers and are typically on the order of seconds. Therefore, the additional $\sim 48$--49~ms introduced by the \ac{PQC} sidecar is operationally compatible with registration and authentication signaling, while still motivating optimization for latency-sensitive deployments.

Quartile analysis highlights distinct distributional behavior across architectures. Native exhibits low-variance distributions, while \ac{TLS} Proxy shows compressed quartiles for \ac{SBI} and container-level metrics but expanded quartiles for Overall Statistics, indicating aggregation variability in the proxy pipeline. By contrast, \ac{PQC} Proxy maintains narrow quartile ranges across all metrics ($IQR \leq 0.2$), demonstrating higher absolute latency with bounded variance---a key property for predictable \ac{QoS}. The separation is clear: \ac{PQC} \ac{SBI} quartiles $[53.9, 54.1]$ do not overlap with Native $[15.0, 15.1]$ or \ac{TLS} Proxy $[10.2, 10.2]$, and the observed $CV<0.4\%$ confirms stable behavior.

\begin{figure}[htbp]
\centering
\includegraphics[width=.48\textwidth]{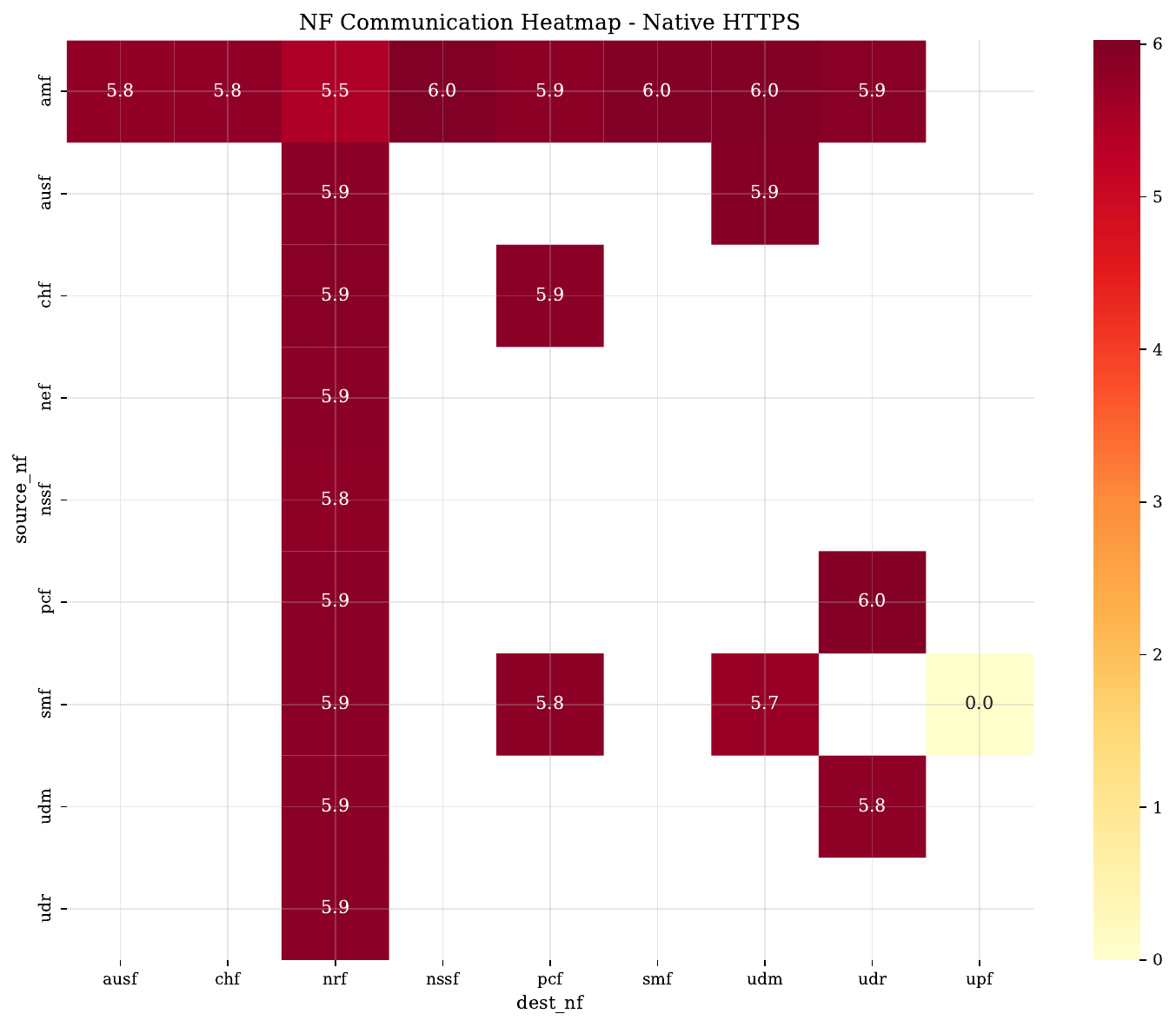}
\caption{Mean latency in the \ac{HTTPS} baseline.}
\label{fig:heatmap_https}
\end{figure}

\begin{figure}[htbp]
\centering
\includegraphics[width=.48\textwidth]{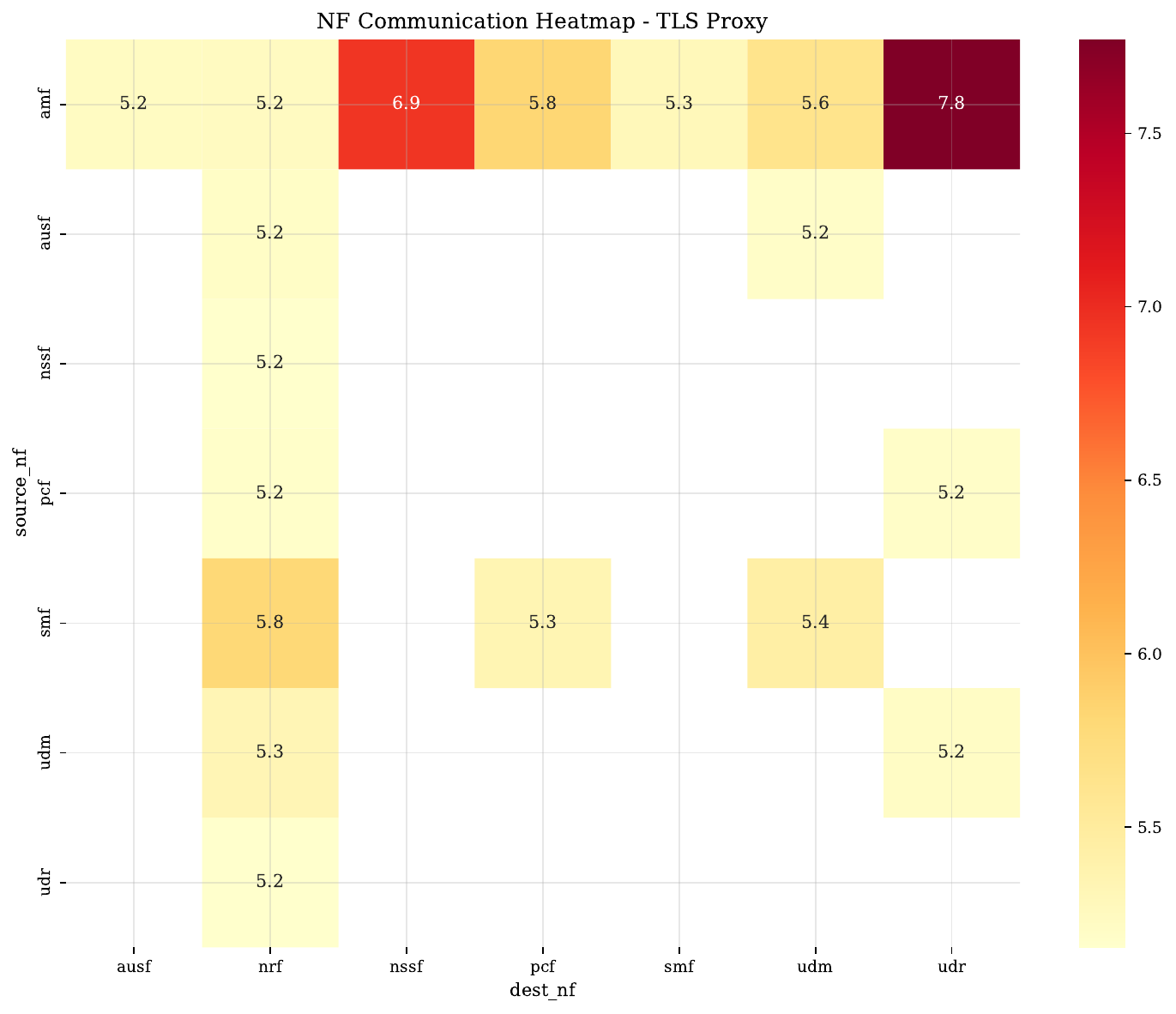}
\caption{Mean latency in the \ac{TLS} Proxy.}
\label{fig:heatmap_tls_proxy}
\end{figure}

\begin{figure}[htbp]
\centering
\includegraphics[width=.48\textwidth]{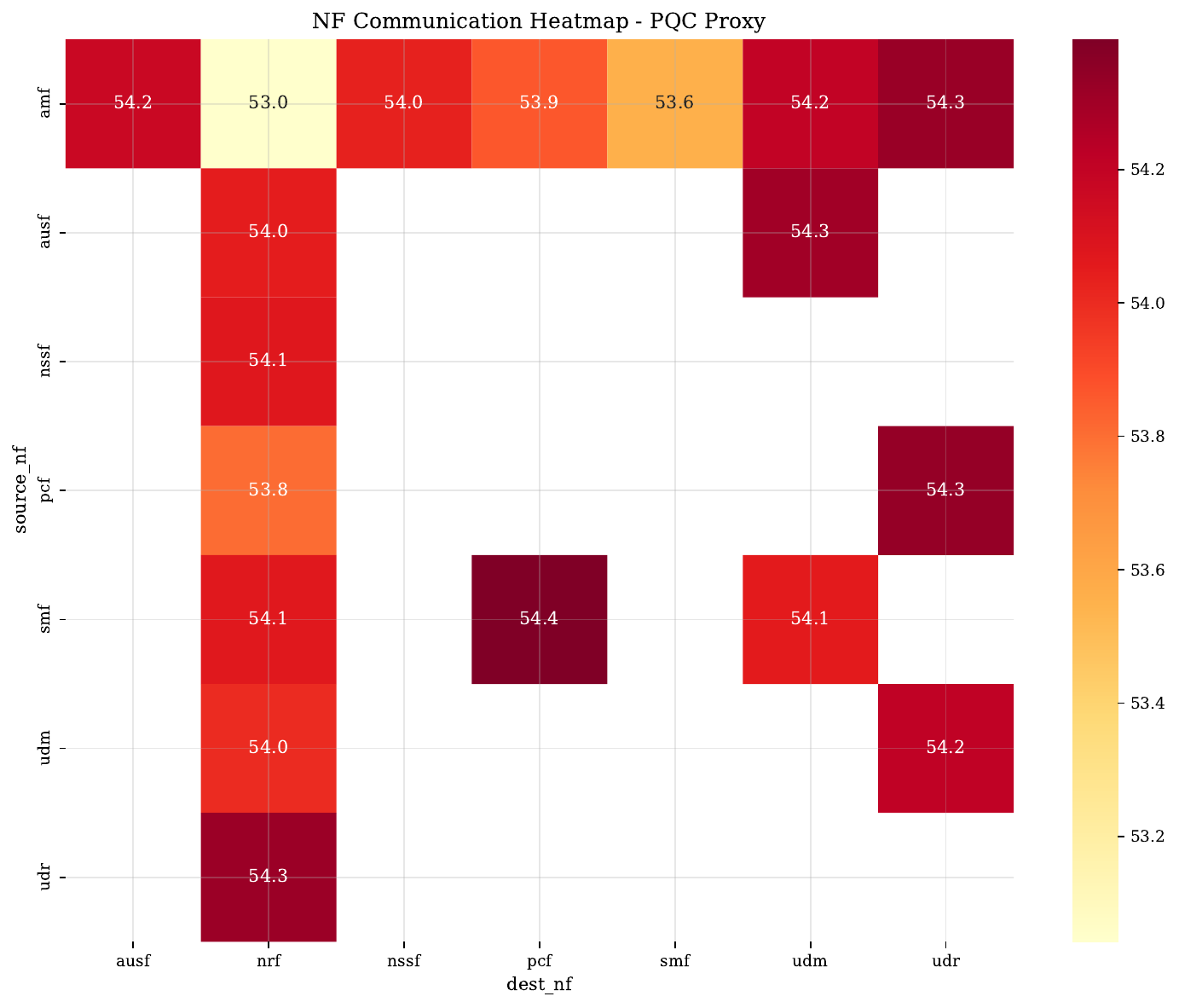}
\caption{Mean latency in the \ac{PQC} Proxy}
\label{fig:heatmap_pqc_proxy}
\end{figure}

Heatmaps further corroborate these trends. Native maintains uniformly low \ac{SBI} latencies (5.5--6.0~ms) across \ac{NF} pairs (Figure~\ref{fig:heatmap_https}), while \ac{TLS} Proxy remains within the same order of magnitude (5.0--7.7~ms) and exhibits stable behavior consistent with optimized connection reuse (Figure~\ref{fig:heatmap_tls_proxy}). \ac{PQC} Proxy exhibits a nearly uniform increase to 53.0--54.4~ms across all \ac{NF} pairs (Figure~\ref{fig:heatmap_pqc_proxy}), indicating deterministic overhead largely independent of the communication pattern and dominated by ML-\ac{KEM}-768 encapsulation/decapsulation and ML-\ac{DSA} processing.

\begin{figure}[htbp]
\centering
\includegraphics[width=.48\textwidth]{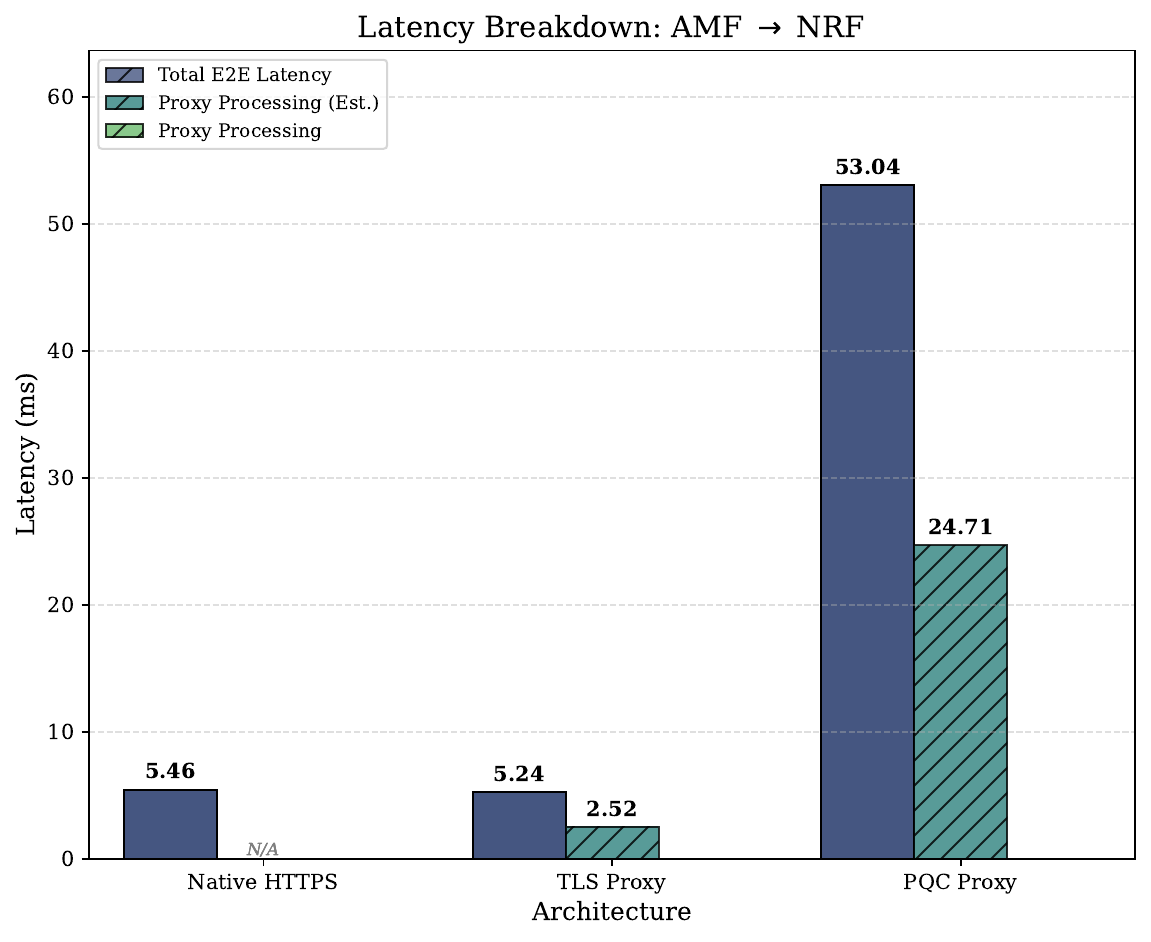}
\caption{Processing latency - \ac{AMF} to \ac{NRF}}
\label{fig:pqc_proxy_latency_amf}
\end{figure}

\begin{figure}[ht]
\centering
\includegraphics[width=.48\textwidth]{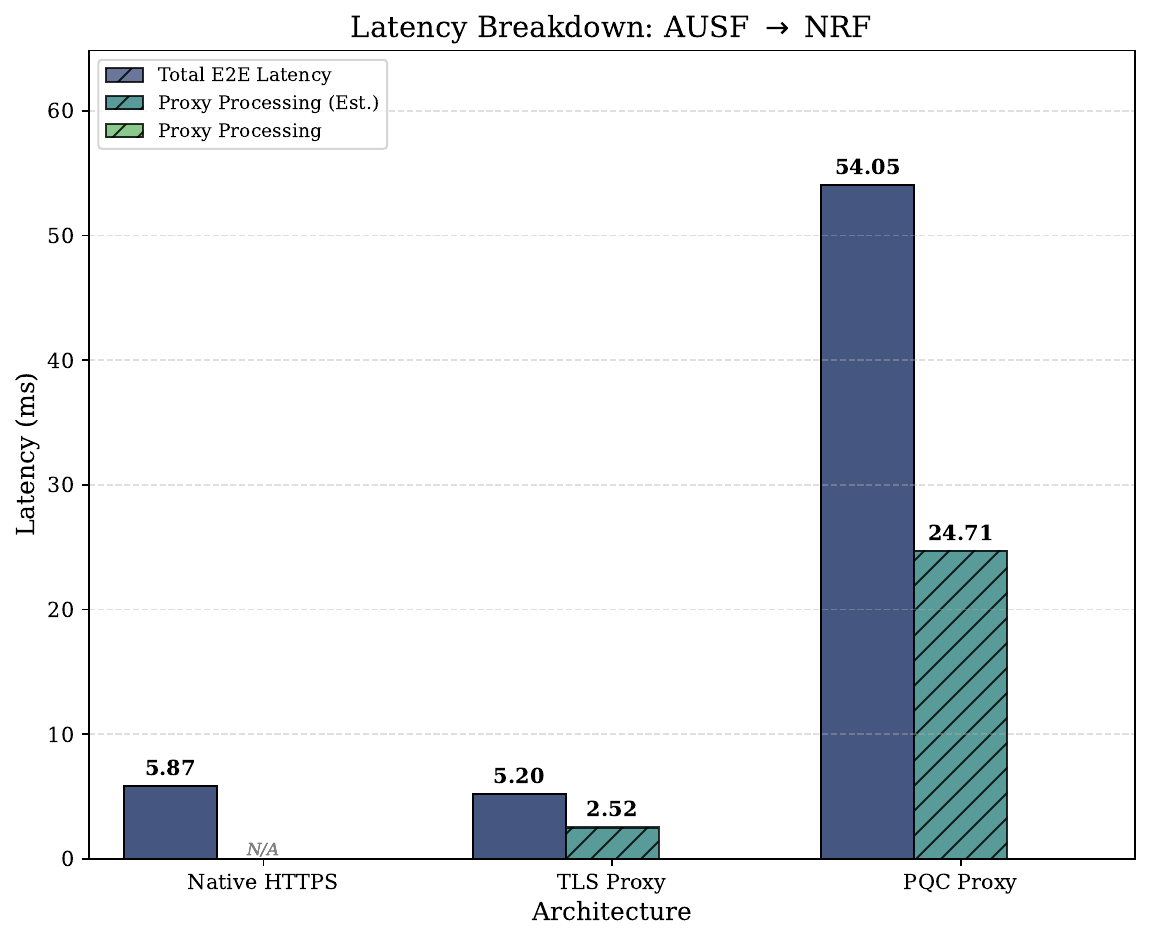}
\caption{Processing latency - \ac{AUSF} to \ac{NRF}}
\label{fig:pqc_proxy_latency_ausf}
\end{figure}

\begin{figure}[ht]
\centering
\includegraphics[width=.48\textwidth]{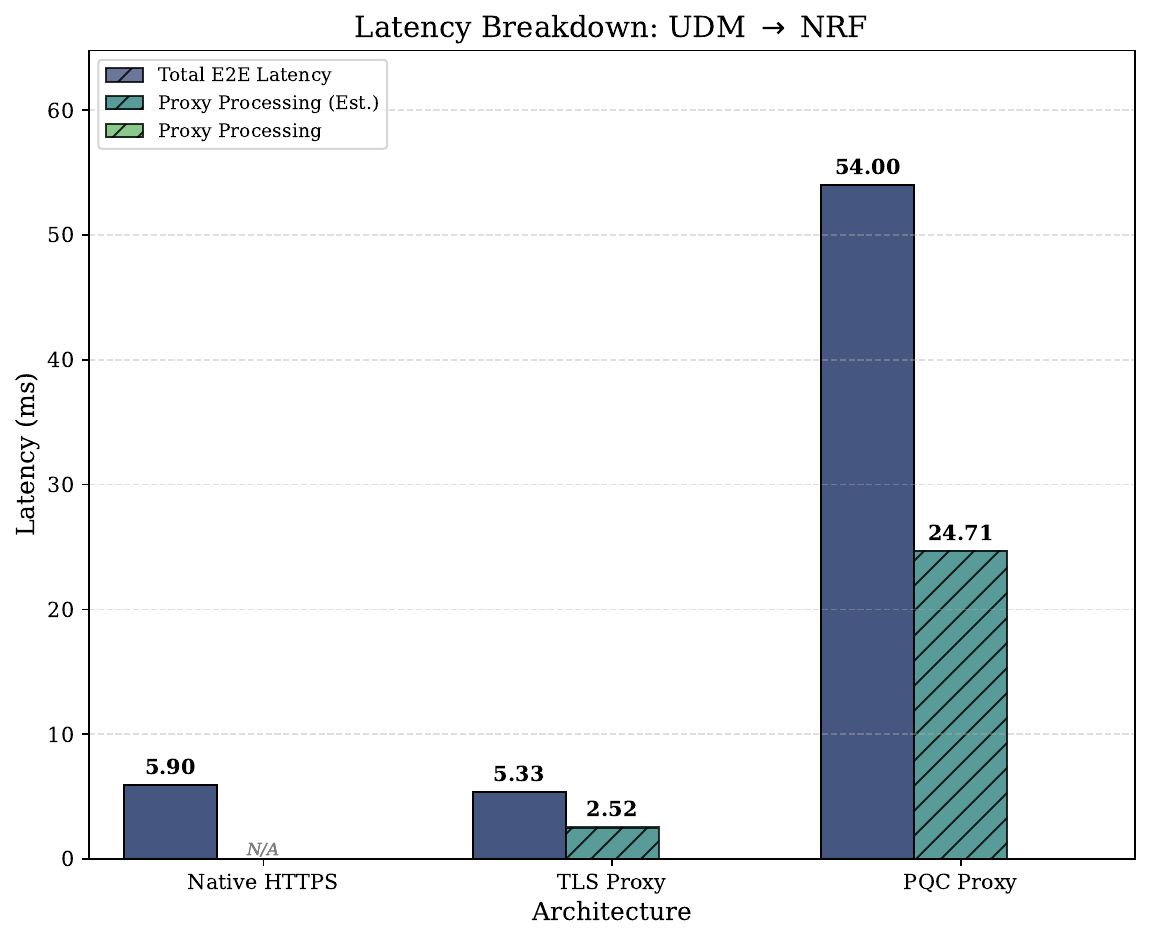}
\caption{Processing latency - \ac{UDM} to \ac{NRF}}
\label{fig:pqc_proxy_latency_udm}
\end{figure}

Detailed latency breakdown for representative \ac{NRF}-centric exchanges (Figures~\ref{fig:pqc_proxy_latency_amf}--\ref{fig:pqc_proxy_latency_udm}) shows the same pattern: Native remains near 5--6~ms, \ac{TLS} Proxy remains near 5~ms with a small proxy processing component (2.43~ms), and \ac{PQC} Proxy increases to 53--54~ms. The proxy processing estimate is stable (24.71~ms) across evaluated pairs, indicating deterministic overhead consistent with encapsulation/decapsulation dominance. The proxy processing estimate is computed inside the sidecar as $\Delta t = t_{\text{egress}} - t_{\text{ingress}}$, capturing cryptographic encapsulation/decapsulation and certificate validation time within the proxy/wrappers, independent of \ac{NF} application processing.


\begin{figure}[htbp]
\centering
\includegraphics[width=.48\textwidth]{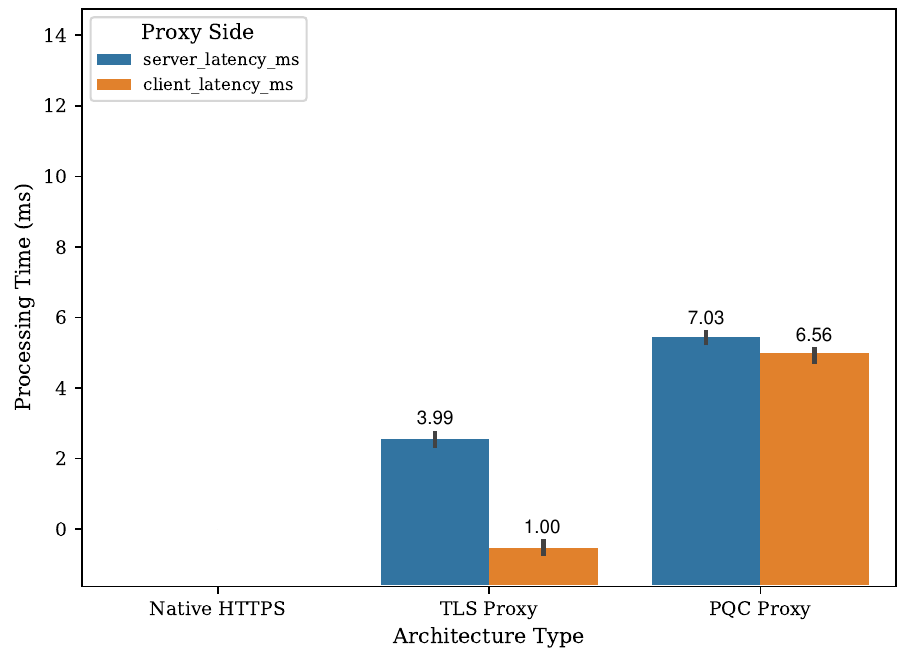}
\caption{PQC latency operations}
\label{fig:pqc_operations_detail}
\end{figure}

The proxy processing breakdown (Figure~\ref{fig:pqc_operations_detail}) reveals asymmetric costs for \ac{TLS} Proxy (server-side 3.99~ms vs.\ client-side 1.00~ms), consistent with server-dominated certificate validation and private-key operations. For \ac{PQC} Proxy, server-side (7.03~ms) and client-side (6.56~ms) costs become near-symmetric (13.59~ms total), indicating that encapsulation and decapsulation impose comparable computational burdens and contribute directly to end-to-end latency. Overall \ac{PQC} overhead comprises (i) ML-\ac{KEM} encapsulation/decapsulation, (ii) ML-\ac{DSA} signature generation/verification, and (iii) proxy-side certificate-chain validation.

Detailed analysis indicates that higher security levels in the \ac{CA} (e.g., ML-\ac{DSA}-87) directly increase server-side validation latency. To quantify the impact of security margins on operational latency, we evaluated three \ac{CA} parameter levels within the \ac{PQC} Proxy configuration.

\begin{table}
\centering
\caption{Impact of \ac{CA} security level on server-side validation latency (PQC Proxy).}
\begin{tblr}{
  width = \linewidth,
  colspec = {Q[248]Q[240]Q[191]Q[242]},
  cell{1}{1} = {font=\bfseries},
  cell{1}{4} = {font=\bfseries},
  row{1} = {Matisse,fg=white,c,m},
  row{2-4} = {m},
  hlines,
  vlines,
}
CA Algorithm & {\textbf{Server Proxy}\\\textbf{(ms)}} & {\textbf{Total SBI}\\\textbf{Median (ms)}} & \textbf{$\Delta$ vs ML-\ac{DSA}-44 (ms)}\\
ML-\ac{DSA}-44 & 6.48 & 53.72 & -\\
ML-\ac{DSA}-65 & 7.03 & 54.05 & +0.33\\
ML-\ac{DSA}-87 & 8.21 & 55.12 & +1.40
\end{tblr}
\label{tab:ca_impact}
\end{table}

As shown in Table~\ref{tab:ca_impact}, increasing the \ac{CA} security level yields a monotonic rise in server-side validation latency: the transition to ML-\ac{DSA}-87 adds +1.40~ms to the median total \ac{SBI} latency relative to ML-\ac{DSA}-44. While this variation is smaller than the dominant PQC overhead, it is repeatable and measurable, confirming certificate-chain validation as a tunable contributor to end-to-end delay. Consequently, \ac{CA} parameter selection becomes a concrete deployment lever for balancing post-quantum security margin and control-plane latency constraints.

We assume a trusted key-generation and distribution service; operational deployments should harden this trust anchor using \acp{HSM} and audited, least-privilege key management to reduce single-point-of-compromise risk.

From a deployment perspective, the results indicate that \ac{PQC}-sidecar integration remains feasible for control-plane operations with moderate latency tolerance, while authentication-intensive \ac{NF} interactions may require resource scaling and/or optimized certificate validation strategies. Lower \ac{CA} security levels provide measurable latency reduction, highlighting a tunable trade-off between cryptographic strength and operational performance.

\section{Conclusion}\label{sec:conclusion}

This work addressed the practical challenge of securing cloud-native \ac{5G} core signaling against future quantum threats, including harvest-now-decrypt-later and active \ac{MITM} attacks, while preserving compatibility with existing \acp{NF} implementations. To bridge the gap between \ac{NIST} \ac{PQC} standardization and deployable telecommunications security, we proposed and implemented a sidecar proxy architecture that transparently provides post-quantum protection for \ac{SBI} communication without modifying legacy \ac{NF} codebases, enabling crypto-agility through a modular security layer. 

Using free5GC, we experimentally compared three scenarios, Native \ac{HTTPS}, \ac{TLS} Sidecar (Proxy/Wrapper), and \ac{PQC} Sidecar (Proxy/Wrapper), under a controlled \ac{SBI} workload centered on \ac{NF} registration, discovery, and heartbeat exchanges. The results show that the \ac{PQC} sidecar increases the end-to-end \ac{SBI} latency to approximately 54 ms, corresponding to an additional 48--49 ms compared to classical \ac{TLS}. However, the overhead remains deterministic and largely uniform across \ac{NF} pairs, with stable distributions that support predictable capacity planning. These findings indicate that transparent \ac{PQC} insertion via sidecars is operationally viable for control plane signaling with moderate latency tolerance and provides a concrete, immediately applicable migration path that preserves operator investments in the current \ac{5G} infrastructure.

Our evaluation focused on control-plane traffic and isolated \ac{SBI} interactions, leaving user-plane protection outside the current scope. The reliance on a trusted key generation and distribution service introduces a critical dependency that must be hardened for production deployments. 

Future work should therefore extend \ac{PQC} coverage to user-plane paths, evaluate hardware acceleration and certificate-validation optimizations for ML-\ac{KEM} and ML-\ac{DSA}, investigate hybrid classical and post-quantum configurations during transition periods, and validate the approach under denser operational regimes such as massive \ac{IoT}, complemented by broader security analysis of the sidecar design under varied threat models.


Overall, by demonstrating a non-disruptive integration approach and quantifying its deterministic performance cost, this study provides a practical foundation for quantum-resistant \ac{5G} core deployments as the ecosystem advances toward \ac{6G}. In this sense, the proposed sidecar pattern translates \ac{NIST}-standardized \ac{PQC} into actionable deployment guidance for operators, enabling incremental adoption of quantum-resilient \ac{SBA} signaling without invasive architectural changes.

\section*{Acknowledgment}

The authors thank FAPEMIG (Grant \#APQ00923-24), FAPESP MCTIC/CGI Research project 2018/23097-3 - SFI2 - Slicing Future Internet Infrastructures, and Fundação para a Ciência e Tecnologia (FCT) within the R\&D Unit Project Scope UID/00319/2025 - Centro ALGORITMI (ALGORITMI/UM) for supporting this work.



\bibliographystyle{IEEEtran}
\bibliography{example}
%

\end{document}